\documentclass[reqno,12pt]{amsart}

\usepackage{amsthm,amsmath,amssymb,amsfonts}
\usepackage{fullpage}

\usepackage{graphicx}
\usepackage{float}
\usepackage{subfig}

\begin{document}

\newtheorem{thm}{Theorem}[section]
\newtheorem{cor}{Corollary}[section]
\newtheorem{lem}{Lemma}[section]
\newtheorem{prop}{Proposition}[section]
\newtheorem{rem}{Remark}[section]

\def\Ref#1{Ref.~\cite{#1}}

\def\const{\text{const.}}
\def\Rnum{{\mathbb R}}
\def\sgn{{\rm sgn}}
\def\i{{\rm i}}
\def\sech{{\rm sech}}

\def\Dop{{\mathcal{D}}}
\def\X{{\rm X}}
\def\Y{{\rm Y}}
\def\pr{{\rm pr}}

\def\Esp{\mathcal{E}}

\def\c{\beta}
\def\b{\alpha}
\def\d{\gamma}
\def\e{\kappa}
\def\h{\tilde h}

\numberwithin{equation}{section}

\allowdisplaybreaks[3]

\title{Symmetries, conservation laws, and generalized travelling waves for a forced Ostrovsky equation}

\author{S.C. Anco${}^1$, M.L. Gandarias${}^2$
\\${}^1$Brock University, St Catharines Canada
\\${}^2$Cadiz University, Cadiz  Spain}

\date{}

\begin{abstract}
Ostrovsky's equation with time- and space- dependent forcing is studied. 
This equation is model for long waves in a rotating fluid with a non-constant depth (topography). 
A classification of Lie point symmetries and low-order conservation laws is presented.
Generalized travelling wave solutions are obtained through symmetry reduction. 
These solutions exhibit a wave profile that is stationary in a moving reference frame
whose speed can be constant, accelerating, or decelerating. 
\end{abstract}


\maketitle

\section{Introduction}

In shallow water theory, 
waves $u(x,t)$ propagating in one spatial dimension are described by 
the well-known Korteweg-de Vries equation (KdV), 
$u_t + uu_x + u_{xxx} =0$. 
The derivation of this equation omits two important physical effects that 
are relevant for shallow ocean waves: 
non-constant water depth, due to bottom topography; 
Coriolis force, due to rotation of the earth. 

The Coriolis effect is modelled by the addition of a term $\c u$, 
as shown by Ostrovsky \cite{Ost} (see also \Ref{Leo,Shr}), 
while the effect of bottom topography is modelled by the addition of a term $h_x(x,t)$
(see e.g. \Ref{Pel,GotGri}). 
These effects lead to the forced Ostrovsky equation 
\begin{equation}\label{FOstrov}
(u_t+uu_x + \b u_{xxx})_x= \c u + h_x
\end{equation}
for the wave amplitude $u(x,t)$, where $\b,\c$ are non-zero constants 
and $h(x,t)$ is a non-constant function of $(x,t)$ which describes the topography. 
Time dependence in $h_x$ models physical situations 
in which the topography is dynamical, 
for instance a submarine landslide or earthquake. 

Very little study has been done on this equation \eqref{FOstrov}. 
The present paper is devoted to studying some basic aspects:  
variational structure, symmetries, conservation laws, and exact invariant solutions. 

Similar work on the KdV equation with forcing and on the unforced Ostrovsky equation
appears in \Ref{GanBru2012,GanKha,ChoIvaLiu}. 
Work on the Cauchy problem as well as on generalizations and reductions of 
the unforced Ostrovsky equation can be found in
\Ref{LevLiu,GriHeOst,GriHek,GilGriSte,NikSteChi,Ste,GriHelJoh,GanBru2010,GanBru2011,KarRazBis}.

In section~\ref{sec:prelim},
the Hamiltonian and Lagrangian formulation of equation \eqref{FOstrov}
is presented.
In section~\ref{sec:symmconslaws},
a classification of Lie point symmetries and Noether conservation laws
is obtained by using the standard methods in \Ref{Olv-book,BCA-book,Anc-review}
(see also \cite{AncBlu97,AncBlu02,Wol}.)
The symmetries comprise a time-dependent Galilean boost 
and a transformation to a moving/shifted reference frame, 
in which the speed can be constant, accelerating, or decelerating. 
These symmetries are shown to be variational 
and yield a conserved energy and a conserved momentum. 

In section~\ref{sec:solns}, 
a plethora of exact invariant solutions are obtained by symmetry reduction
along with the application of a homogeneous balance method. 
The physical meaning of the reductions and the solutions are discussed. 
In particular, the solutions arising from the Galilean symmetry 
describe generalized travelling waves whose profile is stationary in a reference frame 
moving with arbitrary (time-dependent) speed. 
These solutions include a rational solitary wave, a sech-squared solitary wave, 
oscillatory solutions, and static solutions. 

Finally, some concluding remarks are made in section~\ref{sec:remarks}.

\section{Variational structure}\label{sec:prelim}

A Hamiltonian structure exists by introducing a potential $v$ through $u=v_x$,
which allows the equation to be expressed in the evolution form
$u_t +u u_x + \b u_{xxx} =\c v + h$.
The Hamiltonian structure is then given by
\begin{equation}
u_t = -u u_x - \b u_{xxx} +\c v + h
= D_x(\delta H/\delta u)
\end{equation}
where
\begin{equation}
H = \int \big( \tfrac{1}{2}\b u_x{}^2  -\tfrac{1}{6}u^3 -\tfrac{1}{2}\c v^2 + H u\big)\,dx,
\quad
H_x = h . 
\end{equation}

In terms of the potential, the equation takes the form
\begin{equation}\label{FOstrov.pot}
v_{tx}+ v_x v_{xx}+ \b v_{xxxx} -\c v  - h =0
\end{equation}
which possesses a Lagrangian 
\begin{equation}\label{Lagr}
L = -\tfrac{1}{2}v_t v_x -\tfrac{1}{6}v_x{}^3 + \tfrac{1}{2}\b v_{xx}{}^2 - \tfrac{1}{2}\c v^2 + H v_x . 
\end{equation}

Interestingly, it is possible to introduce a further potential via $v=w_x$,
which gives a corresponding 2nd-layer potential equation
\begin{equation}\label{gOstrov.pot.2nd}
w_{tx} + F(w_{xx}) +\b w_{xxxx} -\c w - H =0
\end{equation}
with neither a Hamiltonian nor Lagrangian structure in terms of $w$.

\section{Lie point symmetries and Noether conservation laws}\label{sec:symmconslaws}

Lie point transformations acting on the independent and dependent variables of
the forced Ostrovsky potential equation \eqref{FOstrov.pot}
are generated by vector fields
\begin{equation}\label{symm}
\X =\tau (t,x,v) \partial_t  + \xi(t,x,v)\partial_x  +  \eta (t,x,v) \partial_v .
\end{equation}
The resulting transformation group will be a point symmetry iff
\begin{equation} \label{deteqn}
\pr^{(4)}\X\left(v_{tx}+ v_x v_{xx}+\b v_{xxxx} -\c v -h\right)|_\Esp =0
\end{equation}
where $\pr^{(4)}\X$ is the fourth prolongation of the vector field \eqref{symm},
and $\Esp$ denotes the solution space of equation \eqref{FOstrov}. 
This determining equation \eqref{deteqn} splits with respect to derivatives of $v$,
yielding an overdetermined system of equations for
$\tau(t,x,v)$, $\xi(t,x,v)$, $\eta(t,x,v)$,
together with $h(x,t)$ such that $h_x\not\equiv 0$.

\begin{thm}\label{thm:pointsymms}
(i) The point symmetries admitted by
the forced Ostrovsky potential equation \eqref{FOstrov.pot} 
are generated by:
\begin{align}
\X_{1} = & \partial_t +c(t)\partial_x +\big( c'(t)(x -\smallint c(t)\,dt) + \tfrac{1}{\c}c''(t) - \tfrac{1}{\c} h_0'(t) \big)\partial_v ,
\label{t-trans}
\\
& h(x,t)=  h_1(\chi) - \c c(t)\chi + h_0(t),
\quad
\chi = x-\smallint c(t)\,dt ;
\label{t-trans.h}
\\
&\nonumber\\
\X_{2}= &  e^{-\frac{2}{\c}\smallint h_2(t)\,dt} \partial_x + e^{-\frac{2}{\c}\smallint h_2(t)\,dt}(\tfrac{4}{\c^3} h_2(t)^2 -\tfrac{2}{\c^2} h_2'(t) -\tfrac{1}{\c} h_1(t) -\tfrac{2}{\c}h_2(t) x)\partial_v ,
\label{dil-x-trans-shift}
\\
& h(x,t)= h_2(t) x^2 + h_1(t) x + h_0(t) . 
\label{dil-x-trans-shift.h}
\end{align}
(ii) These symmetries comprise a two-dimensional solvable algebra 
in the case 
\begin{equation}\label{h.quadratic}
h(x,t)= \tilde h_2 x^2 + \tilde h_1(t) x + \tilde h_0(t) . 
\end{equation}
Their commutator is given by $[\X_{1},\X_{2}]= \tfrac{2h_2}{\c}\X_{2}$. 
In terms of $\tilde h_2$, $\tilde h_1(t)$, $\tilde h_0(t)$, 
expressions \eqref{dil-x-trans-shift.h} and \eqref{t-trans.h} are respectively given by 
\begin{equation}\label{X2.overlap}
h_2(t) = \h_2,
\quad
h_1(t) =\h_1(t),
\quad
h_0(t) =\h_0(t)
\end{equation}
and
\begin{equation}\label{X1.overlap}
\begin{gathered}
h_1(\chi)=\h_2\chi^2,
\quad
c(t) =-\big( \tfrac{1}{\c}e^{-\frac{2\h_2}{\c}t}\smallint e^{\frac{2\h_2}{\c}t}\h_1(t)\,dt \big)',
\\
h_0(t)=\h_0(t) -\h_2 (\smallint c(t)\,dt)^2 - \c c(t)\smallint c(t)\,dt .
\end{gathered}
\end{equation}
\end{thm}

Prolongation of the vector fields \eqref{t-trans} and \eqref{dil-x-trans-shift}
to $u(x,t)$ gives the symmetry generators 
\begin{align}
\X_{1} = & \partial_t +c(t)\partial_x + c'(t) \partial_u , 
\label{t-trans.u}
\\
\X_{2}= &  e^{-\frac{2}{\c}\smallint h_2(t)\,dt} \partial_x - \tfrac{2h_2(t)}{\c} e^{-\frac{2}{\c}\smallint h_2(t)\,dt}\partial_u ,
\label{dil-x-trans-shift.u}
\end{align}
both of which are local. 
These symmetries generate the respective transformation groups
\begin{equation}\label{t-trans.transformation}
t\to t +\epsilon,
\quad
x\to x + \epsilon\chi(t+\epsilon),
\quad
u\to u + \epsilon c(t+\epsilon), 
\end{equation}
and 
\begin{equation}\label{dil-x-trans-shift.transformation}
x\to x +\epsilon e^{-\frac{2}{\c}\smallint h_2(t)\,dt},
\quad
u\to u + \epsilon \big( e^{-\frac{2}{\c}\smallint h_2(t)\,dt} \big)',
\end{equation}
with parameter $\epsilon\in\Rnum$. 

Symmetry \eqref{t-trans.transformation} 
describes an accelerated Galilean boost with speed $c(t)$. 
When the speed is constant, $c'(t)=0$, 
this symmetry reduces to an ordinary Galilean boost. 
It becomes a time translation when the speed vanishes, $c(t)\equiv 0$,
which corresponds to the case of spatial forcing, $h=h_1(x)$. 

Symmetry \eqref{dil-x-trans-shift.transformation} describes 
a transformation to a shifted/moving reference frame. 
In the case $h_2(t)\equiv 0$, 
this symmetry reduces to a space translation, 
corresponding to a shift in the (stationary) reference frame.

\subsection{Conservation laws}

Noether's theorem provides a one-to-one correspondence
between variational symmetries and local conservation laws
for the potential equation \eqref{FOstrov.pot}.
A local conservation law is a continuity equation
\begin{equation}\label{conslaw}
(D_t T+D_x \Phi)|_\Esp =0
\end{equation}
where $T$ is the conserved density, and $\Phi$ is the spatial flux,
which are functions of $t$, $x$, $v$, and derivatives of $v$.
Locally trivial conservation laws have the form
$T|_\Esp=D_x\Theta|_\Esp$
being a spatial divergence
and $\Phi|_\Esp = -D_t\Theta|_\Esp$
being a time derivative for some function $\Theta$ of $t$, $x$, $v$, and derivatives of $v$.
Two conservation laws are locally equivalent iff their difference is locally trivial.
Every non-trivial conservation law yields a conserved integral satisfying
\begin{equation}
\frac{d}{dt} \iint_{\Rnum} T\,dx = 0
\end{equation}
for solutions $v(x,t)$ of equation \eqref{FOstrov.pot}
with sufficient asymptotic spatial decay.

The Noether correspondence holds between
variational symmetries in characteristic form $\hat\X|_\Esp = P|_\Esp\partial_v$
and conserved currents $(T,\Phi)|_\Esp$ up to local equivalence.
A symmetry $\hat\X=P\partial_v$ of equation \eqref{FOstrov.pot}
is variational iff
\begin{equation}\label{varsymm.deteqn}
E_v( \pr^{(2)}\X(L) )\equiv 0
\end{equation}
holds off of the solution space,
where $L$ is the Lagrangian \eqref{Lagr}
and $E_v$ denotes the Euler operator (variational derivative) with respect to $v$.
Since the determining equation \eqref{varsymm.deteqn} implies that
a variational symmetry preserves the Lagrangian up to a total divergence,
every variational symmetry preserves the extremals of the Lagrangian
and hence yields a symmetry of the Euler-Lagrange equation \eqref{FOstrov.pot}.
Thus, variational point symmetries can be found by checking
the determining equation \eqref{varsymm.deteqn} for each point symmetry generator
from Theorem~\ref{thm:pointsymms}.
This yields the following result.

\begin{prop}
The Lie point symmetries admitted by 
the forced Ostrovsky potential equation \eqref{FOstrov.pot} 
are variational. 
\end{prop}

The corresponding conservation laws are obtained straightforwardly
by use of a standard homotopy integral formula \cite{BCA-book,Anc-review,AncBlu97} 
(see also \Ref{Olv-book} for a more general version). 

\begin{thm}\label{thm:conslaws}
For the forced Ostrovsky potential equation \eqref{FOstrov.pot}, 
the conservation laws arising from its variational point symmetries
are (up to local equivalence) spanned by the conserved currents:
\\
\begin{equation}\label{conslaw1}
\begin{aligned}
T_{1} = & 
\tfrac{1}{2}\b v_{xx}^2 -\tfrac{1}{6} v_{x}^3 + \tfrac{1}{2}c(t) v_{x}^2  
-\tfrac{1}{2}\c v^2 +(\c c(t)\chi + c'(t) -h_1(\chi)-h_0(t))v , 
\\
X_{1} = & 
\b( v_t +c(t) v_x -c'(t)\chi -\tfrac{1}{\c}(c''(t) -h_0'(t)) )v_{xxx}
-\tfrac{\b}{2}c(t) v_{xx}^2 -\b (v_{tx}-c'(t))v_{xx} + \tfrac{1}{3}c(t) v_{x}^3 \\&
+ \tfrac{1}{2}( v_t -c'(t)\chi - \tfrac{1}{\c}(c''(t) -h_0'(t)) ) v_{x}^2  
+\tfrac{1}{2}v_t^2 -( c'(t)\chi +\tfrac{1}{\c}(c''(t) -h_0'(t)) )v_t \\&
-\tfrac{\c}{2}c(t) v^2 +( \b c(t)^2\chi -c(t) (h_1(\chi)+h_0(t)) )v
-\tfrac{\c}{3} c(t)c'(t)\chi^3 \\&
+\tfrac{1}{2} ( c(t)h_0'(t) + c'(t)h_0(t) - c(t)c''(t) )\chi^2
+\tfrac{1}{\c} h_0(t)(c''(t)-h_0'(t))\chi \\&
+c'(t) \smallint \chi h_1(\chi)\,d\chi
+\tfrac{1}{\c}(c''(t)-h_0'(t))\smallint h_1(\chi)\,d\chi
-\tfrac{\c}{3}c'(t)c(t)(\smallint c(t)\,dt)^3 \\&
-\tfrac{1}{2}(c'(t)h_0(t) +c(t)h_0'(t) - c(t)c''(t))(\smallint c(t)\,dt)^2
+\tfrac{1}{\c}h_0(t)(c''(t)-h_0'(t))\smallint c(t)\,dt ,
\end{aligned}
\end{equation}
holding for $h(x,t)= h_1(\chi) - \c c(t)\chi + h_0(t)$; 
\begin{equation}\label{conslaw2}
\begin{aligned}
T_{2} = & \big( 
\tfrac{1}{2}v_x^2 -\tfrac{2}{\c}h_2(t) v \big) e^{-\frac{2}{\c}  \smallint h_2(t)\,dt} , 
\\
X_{2} = & \big( 
( v_x +\tfrac{2}{\c}h_2(t) x -\tfrac{4}{\c^3} h_2(t)^2 -\tfrac{2}{\c^2} h_2'(t) -\tfrac{1}{\c}h_1(t) ) \b v_{xxx}
-\tfrac{1}{2} \b v_{xx}^2 -\tfrac{2 \b}{\c}h_2(t) v_{xx} +\tfrac{1}{3}v_x^3  \\&
+( \tfrac{1}{\c}h_2(t) x+\tfrac{1}{2\c} h_1(t) -\tfrac{2}{\c^3} h_2(t)^2 -\tfrac{1}{\c^2} h_2'(t) ) v_x^2 
+( \tfrac{2}{\c}h_2(t) x +\tfrac{1}{\c} h_1(t) -\tfrac{4}{\c^3}  h_2(t)^2 -\tfrac{2}{\c^2} h_2'(t) ) v_t \\&
-\tfrac{\c}{2}v^2 
-(h_2(t) x^2 +h_1(t) x +h_0(t)) v 
-\tfrac{1}{2 \c}h_2(t)^2 x^4 
+h_2(t)(\tfrac{4}{3 \c^3} h_2(t)^2 -\tfrac{2}{3 \c^2} h_2'(t) -\tfrac{1}{\c} h_1(t)) x^3  \\&
-( \tfrac{1}{\c} h_0(t)h_2(t) +\tfrac{4}{2\c^3} h_1(t) h_2(t)^2 -\tfrac{1}{2\c} h_1(t)^2 -\tfrac{1}{\c^2} h_1(t)h_2'(t) )x^2 \\&
+h_0(t)( \tfrac{4}{\c^3} h_2(t)^2 -\tfrac{2}{\c^2}  h_2'(t) -\tfrac{1}{\c} h_1(t) )x 
\big) e^{-\frac{2}{\c} \smallint h_2(t)\,dt} ,
\end{aligned}
\end{equation}
holding for $h(x,t)=h_2(t) x^2 + h_1(t) x + h_0(t)$. 
\end{thm}

The resulting conserved integrals consist of 
\begin{equation}
E = \int_{\Rnum} \big( \tfrac{1}{2}\b v_{xx}^2 -\tfrac{1}{6} v_{x}^3 + \tfrac{1}{2}c(t) v_{x}^2  
-\tfrac{1}{2}\c v^2 -(h_1(\chi) -\c c(t)\chi - c'(t)  +h_0(t))v \big)\,dx
\end{equation}
which is energy, 
and 
\begin{equation}
P = \int_{\Rnum} \big( \tfrac{1}{2}v_x^2 -\tfrac{2}{\c}h_2(t) v \big) e^{-\frac{2}{\c}  \smallint h_2(t)\,dt} \,dx
\end{equation}
which is momentum. 
These quantities are generalizations of the well-known 
energy and momentum quantities \cite{LevLiu} for the unforced Ostrovsky potential equation, 
as seen by taking $h\equiv 0$. 

Going back to the original variable $u(x,t)$, 
it is straightforward to see that both 
the energy and momentum conservation laws \eqref{conslaw1}--\eqref{conslaw2}
are nonlocal. 
The energy will have a definite sign if 
$\b<0$, $\c>0$, $c(t)\leq 0$, and $h_1(\chi) -\c c(t)\chi - c'(t)  +h_0(t)\equiv 0$. 
Observe that the last condition holds iff 
$c$ is constant, $h_1(\chi) =\c c\chi$, and $h_0(t)\equiv 0$,
which implies $h\equiv 0$ whereby the forcing vanishes. 
The momentum will have a definite sign if $h_2(t)=0$. 
This implies $h_x = h_1(t)$, namely the forcing is spatially uniform. 

It is useful to remark that the forced Ostrovsky equation itself gives a conserved integral
\begin{equation}\label{mass}
\int_{\Rnum} u \,dx = h(+\infty,t) - h(-\infty,t)
\end{equation}
where the boundary terms refer to $x\to\pm\infty$. 
This quantity \eqref{mass} can be interpreted as the total mass of water 
or as the mean velocity of the water.

\section{Exact invariant solutions}\label{sec:solns}

The Lie point symmetries in Theorem~\ref{thm:pointsymms}
can be used to obtain exact invariant solutions of 
the forced Ostrovsky equation \eqref{FOstrov}
by the well-known symmetry reduction method (see \Ref{Olv-book,BCA-book,BA-book}). 
Three different symmetry reductions can be considered:
$\X_1$; $\X_2$; $\X_3= \X_1 +\mu \X_2$, where $\mu$ is an arbitrary constant.
For each reduction, 
the ODE describing invariant solutions will be derived, 
and explicit solutions of the ODE will be obtained 
by the method of homogeneous balance. 
The resulting solutions consist of 
a plethora of generalized travelling waves 
from $\X_1$ (time-dependent Galilean boosts); 
a quadratic polynomial in $x$ with time-dependent coefficients
from $\X_2$ (time-dependent reference frame shifts);
and cubic polynomial travelling waves 
from $\X_3$.

\subsection{Reduction under time-dependent Galilean boosts}

The first symmetry generator $\X_1$,
which is given by the vector field \eqref{t-trans},  
describes time-dependent Galilean boosts \eqref{t-trans.transformation}. 
Its invariants consist of 
\begin{equation}\label{X1.invs}
\zeta = x -\smallint c(t)\,dt =\chi,
\quad
V=v - c(t)\chi -\tfrac{1}{\c} (c'(t) -h_0(t)) . 
\end{equation}
Hence, an invariant solution for the potential $v$ will have the form 
\begin{equation}\label{X1.inv.soln}
v= V(\zeta) +c(t)\zeta +\tfrac{1}{\c} (c'(t) -h_0(t)) ,
\end{equation}
which determines a solution 
\begin{equation}\label{X1.u}
u= V'(\chi) +c(t) 
\end{equation}
of the forced Ostrovsky equation \eqref{FOstrov}
with $h_x(x,t)= h_1'(\chi) - \c c(t)$.

The invariant function $V(\zeta)$ describes a generalized travelling wave, 
whose profile is stationary in a reference frame that moves with speed $c(t)$. 
This function satisfies the nonlinear fourth-order ODE 
\begin{equation}\label{ODE.V.X1}
\b V'''' +V'V'' -\c V = h_1(\zeta)
\end{equation}
given by symmetry reduction of the potential equation \eqref{FOstrov.pot}. 

There are two different ways to find solutions $V(\zeta)$. 

The first way is to consider a form for $h_1(\zeta)$ such that 
$V(\zeta)$ and its derivatives have a similar form, 
allowing a cancellation to occur among all of the terms in the ODE \eqref{ODE.V.X1}. 
It is readily apparent that this will work when $h_1(\chi)$ is a polynomial in $\chi$, 
since the terms involving $V$ and derivatives of $V$ in the ODE \eqref{ODE.V.X1} 
appear in a polynomial form. 

\emph{Polynomial solutions}:

Hence, polynomial solutions can be sought,
where the degree is determined by the method of homogeneous balance as follows. 
Consider $V(\zeta) = \zeta^n$ for a positive integer $n\geq 1$. 
Substituting this monomial into the ODE produces terms containing the powers 
$n-4,n,2n-3$. 
The only possible balances between pairs of terms occurs for $n=3$,
and thus 
\begin{equation}
V(\zeta) = c_3\zeta^3 + c_2\zeta^2 + c_1\zeta + c_0
\end{equation}
will be considered, 
where the coefficients $c_3,\ldots,c_0$ are constants. 
The form of the ODE shows that, correspondingly, 
the inhomogeneous term must be a cubic polynomial:
\begin{equation}
h_1(\zeta) = a_3\zeta^3 + a_2\zeta^2 + a_1\zeta + a_0
\end{equation}
with constant coefficients $a_3,\ldots,a_0$. 
Substitution of the polynomials $h_1(\zeta)$ and $V(\zeta)$ into the ODE
leads to a system of algebraic equations given by splitting 
the terms in the ODE with respect to the powers of $\zeta$. 
This system can be solved straightforwardly, 
yielding three solution families: 
\begin{equation}\label{X1.solnfam1}
\begin{aligned}
& c_3 = \tfrac{\c \pm\sqrt{72 a_3 +\c^2}}{36},
\quad
c_2 = \tfrac{a_2}{18 c_3 -\c},
\quad
c_1 = \tfrac{18  a_1 a_3 -4  a_2{}^2 -18\c a_1 c_3 +\c^2 a_1}{18  (6  a_3+\c^2) c_3 -(24 a_3+\c^2)\c },
\\& 
c_0 = \tfrac{
36   a_1 a_2 a_3 -8 a_2{}^3 -108 a_0 a_3{}^2 
+2( a_1 a_2 -21 a_0 a_3) \c^2 -18 c_3 (24  a_0 a_3 -2  a_1 a_2+a_0 \c^2) \c  -a_0\c^4}{18 c_3 (24  a_3+\c^2) \c^2 -(108   a_3{}^2+42  \c^2 a_3+\c^4)\c} 
\end{aligned}
\end{equation}
which is a single solution; 
\begin{equation}\label{X1.solnfam2}
\begin{aligned}
& c_3 = \tfrac{\c}{6},
\quad 
c_2 = \tfrac{a_2}{2\c}, 
\quad 
c_1 = \tfrac{(a_0 +c_0\c)\c}{a_2}, 
\quad 
a_3 = \tfrac{\c^2}{3},
\quad 
a_1 = \tfrac{a_2{}^2}{\c^2} ,
\end{aligned}
\end{equation}
which is a one-parameter ($c_0$) family of solutions; 
\begin{equation}\label{X1.solnfam3}
\begin{aligned}
c_3 = \tfrac{\c}{18}, 
\quad
c_1 = \tfrac{(12  c_2{}^2-3 a_1)}{2\c}, 
\quad
c_0 = \tfrac{(12  c_2{}^3 -3  a_1 c_2-\c a_0)}{\c^2}, 
\quad
a_2 =a_3 = 0 , 
\end{aligned}
\end{equation}
which is a one-parameter ($c_2$) family of solutions.

The corresponding solutions for the potential have the form 
\begin{equation}
v= c_3\chi^3 + c_2\chi^2 + (c_1 +c(t))\chi + c_0+\tfrac{1}{\c} (c'(t) -h_0(t)) 
\end{equation}
where $\chi$ is a generalized travelling wave variable \eqref{X1.invs} 
with speed $c(t)$. 
Through the invariants \eqref{inv.soln} and the relation $u=v_x$, 
three solution families of quadratic polynomials in $\chi$ 
\begin{equation}\label{X1.u.soln}
u = 3c_3\chi^2 + 2c_2\chi + c_1 +c(t), 
\quad
h =  a_3 \chi^3 +  a_2 \chi^2 + (a_1 - \c c(t))\chi +a_0 
\end{equation}
are obtained for the forced Ostrovsky equation \eqref{FOstrov}. 
These solutions physically describe generalized travelling waves, 
which accelerate when $c'(t)>0$ and decelerate when $c'(t)<0$. 
The topography is a cubic polynomial in $\chi$ and contains $c(t)$ in the linear term. 
In particular, $c(t)$ appears a term in the forcing $h_x$. 

A second, different way to find solutions of ODE \eqref{ODE.V.X1} is 
by starting with a specified form for $V(\zeta)$ 
and using the ODE to obtain the corresponding form for $h_1(\zeta)$. 
In principle, 
any function $V(\zeta)$ will determine some corresponding function $h_1(\zeta)$. 
But for $h_1(\zeta)$ to look reasonably simple, 
$V(\zeta)$ should be restricted to have a form such that the terms $V''''$ and $V'V''$ in the ODE 
evaluate to a similar form. 
This method succeeds in yielding several types of generalized travelling wave solutions
with speed $c(t)$. 

\emph{Rational solutions}:

Consider, firstly, a rational solution form 
\begin{equation}
V(\zeta) = c_1 \zeta/(c_0 + \zeta^2)
\end{equation}
where $c_1\neq 0$ and $c_0>0$ are constants. 
A similar form will be sought for 
\begin{equation}
h_1(\zeta) = a_1 \zeta/(c_0 + \zeta^2) + a_2 \zeta/(c_0 + \zeta^2)^2 + a_3 \zeta/(c_0 + \zeta^2)^3 . 
\end{equation}
Substitution of these expressions into the ODE \eqref{ODE.V.X1} 
leads to a system of algebraic equations for their coefficients. 
The solution is given by 
\begin{equation}
c_1=24\b, 
\quad
a_1 = -24\b\c,
\quad
a_2 =0, 
\quad
a_3 = -(a_1/\c){}^2 , 
\end{equation}
which is a one-parameter ($c_0$) family. 
This yields a resulting solution family for $u=v_x$:
\begin{equation}
u = c(t) + \frac{24\b\c(\chi^2 -c_0)}{(\chi^2+c_0)^2},
\quad
h = -\frac{24\b\c \chi}{\chi^2+c_0} -\frac{(24\b)^2\chi}{(\chi^2+c_0)^3},
\quad
\chi = x - \smallint c(t)\,dt .
\end{equation}
These generalized travelling waves will be non-singular for $c_0>0$. 
They describe single-hump heavy tail waves on a background $c(t)$, 
where the topography has the shape of an antisymmetric heavy tail wave. 

\emph{Solitary wave solutions}:

Consider, secondly, a tanh solution form 
\begin{equation}
V(\zeta) = c_1 \tanh(c_0\zeta) 
\end{equation}
where $c_0,c_1\neq 0$ are constants. 
A similar form will be sought for 
\begin{equation}
h_1(\zeta) = a_1 \tanh(c_0\zeta) + a_2 \tanh(c_0\zeta)^2+ a_3 \tanh(c_0\zeta)^3 . 
\end{equation}
Substitution of these expressions into the ODE \eqref{ODE.V.X1} 
gives a system of algebraic equations for their coefficients. 
The solution is given by 
\begin{equation}
c_1=12\b c_0, 
\quad
a_1 = -12(8c_0{}^4\b  +\c)\b,
\quad
a_2 =0, 
\quad
a_3 = 96 c_0{}^5 \b^2 . 
\end{equation}
This a one-parameter ($c_0$) family. 
The resulting solution family for $u=v_x$ has the form 
\begin{equation}
u = c(t) +\frac{12\b c_0{}^2}{\cosh(c_0\chi)^2} , 
\quad
h = -12c_0 \b\Big( \c +\frac{8 c_0{}^4 \b}{\cosh(c_0\chi)^2}\Big)\tanh(c_0\chi) ,
\quad
c_0>0,
\quad
\chi = x - \smallint c(t)\,dt .
\end{equation}
This describes generalized solitary waves on a background $c(t)$,
where the shape of the topography 
either is a shock when $(\c+8 c_0{}^4\b)\b >0$,
or otherwise is a double-peaked shock. 
Note that the shape will be a tanh-cubed expression when 
$c_0{}^4 = -\tfrac{\c}{8\b}>0$. 

\emph{Oscillatory solutions}:

Finally, consider a sine-cosine solution form 
\begin{equation}
V(\zeta) = c_0 + c_1 \cos(\omega\zeta+\phi) 
\end{equation}
where $c_1,\omega\neq 0$ and $c_0,\phi$ are constants. 
A similar form will be sought for 
\begin{equation}
h_1(\zeta) = a_0 + a_1 \cos(w\zeta +\psi) . 
\end{equation}
The ODE \eqref{ODE.V.X1} leads to a system of algebraic equations for their coefficients.
Solving the system yields 
\begin{equation}
a_1 = -\tfrac{1}{2}c_1{}^2\omega^3,
\quad
a_0 = - c_0\c, 
\quad
w= 2\omega,
\quad
\psi = 2\phi, 
\quad
\omega^4 = \c/\b >0, 
\end{equation}
which is a three-parameter ($\phi,c_0,c_1$) family. 
The resulting solution family for $u=v_x$ is given by 
\begin{equation}
u = c(t) -c_1\omega\sin(\omega\chi +\phi) , 
\quad
h = -c_0\c -\tfrac{1}{2}c_1{}^2\omega^3\sin(2\omega\chi +2\phi) , 
\quad
\omega=\sqrt[4]{\tfrac{\c}{\b}}, 
\quad
\chi = x - \smallint c(t)\,dt . 
\end{equation}
These solutions describe oscillatory generalized travelling waves on a background $c(t)$,
where the shape of the topography is oscillating 
with twice the frequency of oscillation of the waves. 

\emph{Static solutions}:

All of the previous solutions continue to hold when the topography is time-independent, $h_t\equiv 0$. 
In this situation, the wave speed vanishes, $c(t)\equiv 0$, 
and the solutions become static.

\subsection{Reduction under time-dependent reference frame shifts}

The second symmetry generator $\X_2$,
which is given by the vector field \eqref{dil-x-trans-shift}, 
describes a transformation to a shifted/moving reference frame \eqref{dil-x-trans-shift.transformation}. 
Its invariants are given by 
\begin{equation}\label{X2.invs}
\zeta = t,
\quad
V=v +\tfrac{1}{\c}h_2(t) x^2  +(\tfrac{1}{\c} h_1(t) +\tfrac{2}{\c^2} h_2'(t) -\tfrac{4}{\c^3} h_2(t)^2) x . 
\end{equation}
Hence, an invariant solution for the potential $v$ will have the form 
\begin{equation}\label{X2.inv.soln}
v= V(\zeta) - \tfrac{1}{\c}h_2(\zeta) x^2  -(\tfrac{1}{\c} h_1(\zeta) +\tfrac{2}{\c^2} h_2'(\zeta) -\tfrac{4}{\c^3} h_2(\zeta)^2) x . 
\end{equation}
Reduction of the potential equation \eqref{FOstrov.pot} is trivial --- namely, 
the invariant function $V(\zeta)$ is arbitrary. 

This determines a solution 
\begin{equation}\label{X2.u}
u= - \tfrac{2}{\c}h_2(t) x  -(\tfrac{1}{\c} h_1(t) +\tfrac{2}{\c^2} h_2'(t) -\tfrac{4}{\c^3} h_2(t)^2) 
\end{equation}
of the forced Ostrovsky equation \eqref{FOstrov}
with $h(x,t)= h_2(t) x^2 + h_1(t) x + h_0(t)$. 
Both the solution, $u$, and the forcing, $h_x$, 
are linear polynomials in $x$ with time-dependent coefficients. 
Observe that the invariant function $V(\zeta)$ does not appear in the solution.

\subsection{Travelling wave reduction}

Now consider the symmetry generator
\begin{equation}\label{symm.reduc}
\begin{aligned}
\X= &  \X_1 + \mu\X_2 \\
= & \partial_t +(c(t) +\mu e^{-\frac{2\h_2}{\c}t})\partial_x 
+\big( (2\mu e^{-\frac{2\h_2}{\c}t}-c'(t))x 
-\mu e^{-\frac{2\h_2}{\c}t}( c(t) + \tfrac{2\h_2}{\c}\smallint c(t)\,dt + \tfrac{4\h_2{}^2}{\c^3} ) \\&
-c(t)^2 -\tfrac{2\h_2}{\c}c(t)\smallint c(t)\,dt -\tfrac{1}{\c}c''(t) +\tfrac{1}{\c} \h_0'(t) \big)\partial_v ,
\quad
\mu\neq 0, 
\end{aligned}
\end{equation}
which is a linear combination of the two Lie point symmetries \eqref{t-trans} and \eqref{dil-x-trans-shift}
admitted when $h(x,t)$ has the form \eqref{h.quadratic}. 
Here,  $c(t)$ is given by expression \eqref{X1.overlap}. 
The invariants of this symmetry can be expressed in the form 
\begin{equation}\label{inv1}
\zeta = x -\smallint\tilde c(t)\,dt,
\quad
\tilde c(t) = \begin{cases}
c(t) -\tfrac{\mu\c}{2\h_2} e^{-\frac{2\h_2}{\c}t},
& \h_2\neq 0 
\\
c(t) +\mu t,
& \h_2= 0 
\end{cases}
\end{equation}
and
\begin{equation}\label{inv2}
V=\begin{cases}
v - \tilde c(t) x - \tfrac{\h_2}{\c}(\smallint\tilde c(t)\,dt)^2 -\tfrac{1}{\c}\tilde c'(t) +\tfrac{1}{\c}\h_0(t), 
& 
\h_2\neq 0 
\\
v + (\mu -\tilde c(t)) x -\tfrac{1}{\c}\tilde c'(t) +\tfrac{1}{\c}\h_0(t), 
&
\h_2= 0 
\end{cases} 
\end{equation}
where $\tilde c(t)$ is a wave speed. 
Hence, an invariant solution will have the form 
\begin{equation}\label{inv.soln}
v=\begin{cases}
V(\zeta) + \tilde c(t) x +\tfrac{\h_2}{\c}(\smallint\tilde c(t)\,dt)^2 +\tfrac{1}{\c}\tilde c'(t) -\tfrac{1}{\c}\h_0(t), 
& \h_2\neq 0
\\
V(\zeta) -(\mu-\tilde c(t)) x +\tfrac{1}{\c}\tilde c'(t) -\tfrac{1}{\c}\h_0(t), 
& \h_2= 0
\end{cases}
\end{equation}
in the two separate cases. 

In the case $h_2\neq 0$, 
the function $V(\zeta)$ satisfies the nonlinear fourth-order ODE 
\begin{equation}\label{ODE.V.case1}
\b V'''' +V'V'' -\c V = \h_2\zeta^2
\end{equation}
given by symmetry reduction of the potential equation \eqref{FOstrov.pot}. 
This reduction can be seen to be equivalent to a special case of 
the time-dependent Galilean boost reduction, 
with $h_1(\zeta) = \h_2\zeta^2$. 
Hence, no new solutions can be obtained for the forced Ostrovsky equation \eqref{FOstrov}. 

In the case $\h_2=0$, 
symmetry reduction of the potential equation \eqref{FOstrov.pot} yields 
the nonlinear fourth-order ODE 
\begin{equation}\label{ODE.V.case2}
\b V'''' +(V' -\mu)V'' -\c V = 0. 
\end{equation}
This ODE has the same form as the travelling wave ODE for the unforced Ostrovsky potential equation, 
where $\mu$ would be the travelling wave speed. 
When $c(t)\not\equiv 0$, the form of $u=v_x= \tilde c(t) + V'(\zeta)$ 
still describes a generalized travelling wave in which the profile 
$V'(\zeta)=V'(x-\smallint \tilde c(t)\,dt)$ 
is stationary in a reference frame that moves with speed $\tilde c(t)$. 
This becomes a constant speed travelling wave in the case $c(t)\equiv 0$, 
where $\tilde c(t) = \mu t$. 

It is well known that an explicit form has not yet been found \cite{GilGriSte,NikSteChi}
for solitary waves or periodic waves of Ostrovsky's equation. 
In particular, 
all of the standard special solution methods,
such as the tanh and cosh methods (or the equivalent $G'/G$ method),
the simplest equation method, and their numerous variants, 
fail to yield any solutions for Ostrovsky's equation. 

Nevertheless, 
the ODE \eqref{ODE.V.case2} possesses a family of exact polynomial solutions,
which can be derived similarly to the polynomial solutions found 
for the invariant ODE under time-dependent Galilean boosts. 
This yields 
\begin{equation}
V(\zeta) = 
\tfrac{\c}{18} \zeta^3 +c_2 \zeta^2 +\tfrac{12 c_2{}^2 -\c \mu}{2\c} \zeta +\tfrac{3 c_2(4c_2{}^2 -\c \mu)}{\c^2} 
\end{equation}
with $\zeta = x -\smallint \tilde c(t)\,dt$ in terms of the wave speed 
$\tilde c(t) = \mu + c(t)$. 
Through expression \eqref{inv2} in the case $\h_2=0$, 
a one-parameter ($c_2$) family of quadratic polynomial solutions $u=v_x$ is obtained
\begin{equation}\label{u.soln.case2}
u = \tilde c(t) + \tfrac{\c}{6} (x-\smallint \tilde c(t)\,dt)^2 +2c_2 (x-\smallint \tilde c(t)\,dt) +\tfrac{3(4 c_2{}^2 -\c \mu)}{2\c}, 
\quad
\h = \h_1(t)x +\h_0(t) 
\end{equation}
for the forced Ostrovsky equation \eqref{FOstrov} 
with a topography that is linear in $x$. 
Here, $\tilde c(t)$ is related to $\h_1(t)$ by 
\begin{equation}
\tilde c(t) =\mu  -\big( \tfrac{1}{\c}e^{-\frac{2\h_2}{\c}t}\smallint e^{\frac{2\h_2}{\c}t}\h_1(t)\,dt \big)' 
\end{equation} 
via expression \eqref{X1.overlap}. 

This solution family physically describes generalized travelling waves 
whose profile is stationary after a time-dependent Galilean boost to a reference frame moving with speed $\tilde c(t)$. 
In the case $\h_1(t)\equiv 0$, the physical speed is constant, $\tilde c = \mu$,
and the underlying symmetry is a translation 
\begin{equation}\label{symm.case2}
t\to t +\epsilon,
\quad
x\to x +\mu \epsilon 
\end{equation}
under which $\zeta = x-\mu t$ is invariant.

\subsection{First integrals from symmetry reduction}

Finally, 
some further remarks on the invariant ODEs \eqref{ODE.V.case2} and \eqref{ODE.V.X1}
will be worthwhile. 
Since each ODE arises from a symmetry reduction of the potential equation \eqref{FOstrov.pot}, 
which possesses a Lagrangian \eqref{Lagr}, 
the corresponding conservation law of the potential equation will similarly reduce 
to a first integral of the ODE. 
This is a consequence of general result for symmetry reduction of Euler-Lagrange PDEs
\cite{Sjo2007,Sjo2009,BokDweZamKarMah}. 
Furthermore, all first integrals that arise from symmetry invariant conservation laws
can be found directly by using the symmetry,
through the general multi-reduction method introduced in \Ref{AncGan}. 

In the case of the travelling wave symmetry \eqref{symm.case2}, 
multi-reduction shows that both the energy and momentum conservation laws \eqref{conslaw1}--\eqref{conslaw2}
are invariant, where $h = \h_0(t)$ (with $\h_2=\h_1\equiv 0$). 
This reduction yields two first integrals, 
which are readily seen to coincide (up to a constant factor). 
The resulting first integral of ODE \eqref{ODE.V.case2} is given by 
\begin{equation}\label{FI.V.case2}
\begin{aligned}
\Psi & = 
\c V' V''' -\tfrac{\c}{2} V''^2 + \tfrac{1}{3\b}  V'^3 -\tfrac{\mu}{2\b} V'^2 -\tfrac{\c}{2} V^2 
=\const .
\end{aligned}
\end{equation}
This first integral \eqref{FI.V.case2} does not involve $\h_0(t)$ 
and therefore it also holds for the travelling wave ODE 
arising from the unforced Ostrovsky equation. 

In contrast, 
for the case of the time-dependent Galilean symmetry \eqref{t-trans.transformation}, 
multi-reduction shows that the only invariant conservation law is the energy \eqref{conslaw1}. 
However, under symmetry reduction, the resulting first integral is trivial. 
Consequently, no non-trivial first integrals are inherited by the invariant ODE \eqref{ODE.V.case1}
when $h_1'\not\equiv 0$.

\section{Concluding remarks}\label{sec:remarks}

The present work has obtained several new results 
for the forced Ostrovsky equation.

Firstly, 
all Lie point symmetries and corresponding Noether conservation laws have been classified. 
Their physical meaning has been described, 
which helps to inform the existence and the properties of 
the symmetries and the conserved integrals.
In particular, the symmetries describe a time-dependent Galilean boost 
and a transformation to a moving/shifted reference frame,
both of which involve an arbitrary (time-dependent) speed. 
These symmetries are variational and yield a conserved energy and a conserved momentum. 

Secondly, as an application of this classification, 
symmetry reduction has been used to derive 
a plethora of exact invariant solutions. 
From the Galilean symmetry, 
generalized travelling wave are obtained 
in which the wave profile is stationary in a reference frame moving with 
arbitrary (time-dependent) speed. 
These solutions include a rational solitary wave, a sech-squared solitary wave, 
oscillatory solutions, and static solutions. 
In the case of constant speed, 
a first integral is obtained for the travelling wave ODE. 
From the moving/shifted reference frame symmetry, 
polynomial solutions in $x$ with coefficients depending on $t$ are obtained. 

These explicit invariant solutions can be expected to be important 
in understanding the asymptotics of 
more general solutions of the forced Ostrovsky equation \eqref{FOstrov}. 

Unlike other nonlinear dispersive wave equations, 
no exact solutions of the Ostrovsky equation (forced and unforced) arise 
from use of the typical special solution methods such as 
the tanh and cosh methods (or the equivalent $G'/G$ method), 
the simplest equation method, and their numerous variants. 
In contrast, 
the present work illustrates the importance of selecting a method that is 
suitably adapted to the form of a given wave equation. 
Namely, the equation should inform what method to use, 
rather than the obverse approach.

\section*{Acknowledgments}
SCA is supported by an NSERC Discovery Grant.
MG is supported by FQM-201 from Junta de Andalucia and C\'adiz University Plan propio.

This paper is dedicated to our colleague Masood Khalique,
in recognition of his retirement 
and long-standing work on exact solutions of nonlinear PDEs.

 \end{document}